# Revealing Controllable Anisotropic Magnetoresistance in Spin-orbit Coupled Antiferromagnet $Sr_2IrO_4$

Chengliang Lu[*], Bin Gao, Haowen Wang, Wei Wang, Songliu Yuan, Shuai Dong[*], and Jun-Ming Liu

Dr. C. L. Lu, H. W. Wang, W. Wang, Prof. S. L. Yuan

School of Physics & Wuhan National High Magnetic Field Center,

Huazhong University of Science and Technology, Wuhan 430074, China

Email: cllu@hust.edu.cn

B. Gao, Prof. S. Dong

School of Physics, Southeast University, Nanjing 211189, China

Email: sdong@seu.edu.cn

Prof. J. –M. Liu

Laboratory of Solid State Microstructures and Innovation Center of Advanced Microstructures,

Nanjing University, Nanjing 210093, China

Institute for Advanced Materials, Hubei Normal University, Huangshi 435001, China

# [Abstract]

Antiferromagnetic spintronics actively introduces new principles of magnetic memory, in which the most fundamental spin-dependent phenomena, i.e. anisotropic magnetoresistance effects, are governed by an antiferromagnet instead of a ferromagnet. A general scenario of the antiferromagnetic anisotropic magnetoresistance effects mainly stems from the magnetocrystalline anisotropy related to spin-orbit coupling. Here we demonstrate magnetic field driven contour rotation of the fourfold anisotropic magnetoresistance in bare antiferromagnetic $Sr_2IrO_4$/$SrTiO_3$ (001) thin films hosting a strong spin-orbit coupling induced $J_{eff}$=1/2 Mott state. Concurrently, an intriguing minimal in the magnetoresistance emerges. Through first principles calculations, the band-gap engineering due to rotation of the Ir isospins is revealed to be responsible for these emergent phenomena, different from the traditional scenario where relatively more conductive state was obtained usually when magnetic field was applied along the magnetic easy axis. Our findings demonstrate a new efficient route, i.e. via the novel $J_{eff}$=1/2 state, to realize controllable anisotropic magnetoresistance in antiferromagnetic materials.

# 1. Introduction

Antiferromagnets have been garnering increasing interest in the spintronics community, due to their intrinsic properties such as zero stray magnetic field, ultrafast spin dynamics, and rigidity against external fields [1-4]. The observation of anisotropic magnetoresistance (AMR) and even memory effect in antiferromagnetic (AFM) materials represent a major step in the field of AFM spintronics in which the antiferromagnet governs the transport instead of just playing a passive supporting role in the traditional ferromagnetic (FM) spintronics [5-8]. Here the fundamentals rely mainly on the magnetocrystalline anisotropy in the antiferromagnets, arising from the relativistic spin-orbit coupling (SOC). Since this anisotropy energy is an even function of ordered spins [9, 10], it is feasible to electrically read out the spin axis of staggered spins in an antiferromagnet through AMR effects which is the magnetotransport counterpart of the anisotropy energy. For instance, as confirmed in several AFM materials, rotating the AFM spin-axis with respect to crystal axes leads to variations in electric conductivity, and as a consequence AMR is evidenced [5-8, 11-13]. Although previous studies have successfully realized AFM-based AMR (AFM-AMR) through various approaches, it remains a great challenge to tailor the AMR in AFM materials, which hinders the recognition of the full merits of antiferromagnets.

Recently, the strong SOC in some AFM iridates is found to be essentially involved in Mottness of 5$d$ electrons, i.e. opening a band-gap by the collaborative effect of strong SOC and moderate Hubbard repulsion, forming a novel $J_{eff}$=1/2 state [14, 15]. Such a $J_{eff}$=1/2 state, with the SOC as its ingredient, provides nontrivial playground to engineer the AFM-AMR, which may lead to unusual AMR phenomena. In addition, the strong SOC in iridates essentially entangle both the spin and orbital momenta, thus giving rise to a $J_{eff}$ =1/2 character of the Ir magnetic moments [16]. This is fundamentally distinct from the situation in most previously studied AFM alloys where the SOC and magnetic moments often have different sources, i.e. the SOC arises from heavy elements while the magnetic moments come from 3$d$ transition metals [2, 3]. Therefore, the iridates with novel $J_{eff}$=1/2 state are leading candidates for comprehensively understanding the physical link between the AMR and magnetocrystalline anisotropy related to strong SOC, and exploring possibly controllable AFM-AMR.

Here we demonstrate a contour rotation of the fourfold AMR in the prototypical $J_{eff}$=1/2 AFM semiconductor $Sr_2IrO_4$ (SIO) thin films grown on (001) $SrTiO_3$ (STO) substrate. Concurrently, an abnormal magnetoresistance (MR) minimal is evidenced. With the help of first-principles calculations, the band-gap engineering due to the rotation of Ir's isospins is revealed to be responsible for these phenomena, as a new route to manipulate the AFM-AMR.

## 2. Results and Discussion

### 2.1. Sample preparation and characterization

Because of IrO$_6$ octahedra rotation ($\alpha$~11.8°) with respect to the $c$ axis, Sr$_2$IrO$_4$ has an expanded tetragonal cell. The Ir isospins, containing remarkable orbital contribution, prefer the AFM order within the $ab$ plane, and show a collective deviation of the isospins from the $a$ axis, defining the isospin canting angle $\phi$~13° [17, 18]. Jackeli et al. theoretically revealed that the angle $\phi$ rigidly tracks the lattice distortion $\alpha$, i.e., $\alpha$~$\phi$ [19]. A net moment was found to exist in each IrO$_2$ layer, arising from the isospin canting [17, 20-23]. By applying small magnetic field within the $ab$ plane, a spin-flip transition can be triggered, resulting in a weak FM phase in Sr$_2$IrO$_4$ [20, 24, 25]. This provides a natural handle to drive the planar AFM orders by external magnetic field in Sr$_2$IrO$_4$. The magnetic structure is schematically shown in **Figure 1**(a).

To preserve the spin-orbit coupled AFM orders, epitaxial Sr$_2$IrO$_4$ thin films with thickness of ~40 nm were grown on (001) STO substrates, considering the good lattice fit between the film and the substrate. The Sr$_2$IrO$_4$ thin films exhibit a layer-by-layer growth mode, evidenced by the 2-dimensional reflection high energy electron diffraction (RHEED) image and intensity oscillations (**Figure S1**). The thin films are of pure phase and high quality, identified by detailed structural characterizations (**Figure S2**) [26]. The x-ray reciprocal space mappings (RSM) shown in Figure 1(b) reveals the coherent growth of the films on the substrates, and the in-plane and out-of-plane lattice parameters are found to be $a/\sqrt{2}= b/\sqrt{2}$=3.905 Å and $c/2$=12.848 Å. A small tensile strain $\varepsilon$ of only ~ 0.46% is detected as expected. For a convenience, here a pseudo-tetragonal lattice expression that has a 45° in-plane rotation with respect to the tetragonal lattice is employed for the films, as schematically shown in Figure 1(c).

**Figure 2**(a) presents resistivity ($\rho$) as a function of $T$ for the film. A semiconducting behavior is seen in the entire $T$ range, resulting from the $J_{eff}$=1/2 Mott state [14]. In the $M(T)$ curves of both field cooling (FC) and zero field cooling (ZFC) sequences shown in Figure 2(b), clear FM phase transition can be identified at $T$=$T_C$ ~230 K, arising from the Ir isospin canting as sketched in Figure

1(a). Note that here the FM transition is sharp and the $T_C$ is close to the value seen in $Sr_2IrO_4$ bulk crystals [20, 24], confirming the high quality of our thin films. Further evidence to the weak FM phase is provided by *M(H)* measurements at various temperatures, and a saturated magnetization is estimated to be $M_s$~0.03 μ$_B$/Ir at *T*=10 K (**Figure S3**), which is slightly smaller than the value in bulk crystals [24]. Accompanying with the field induced isospin-flip transition at *H*~0.2 T, *ρ* shows a quick decrease with *H*, evidencing a significant suppression of the spin-dependent transport scattering effect (Figure S3).

By applying *H* along different directions, two interesting features can be seen in the MR data measured at various temperatures, shown in Figure 2(c) and (d). First, the MR curves of *H*//[100] and *H*//[110] are different from each other, and present an intriguing intercross at high field region, evidencing an unconventional AMR phenomenon in the thin films. Second, accompanying with the intercross, the MR curve of *H*//[110] shows a break-in-slope upon sweeping-up *H* continuously. For instance, with *H*//[110] at *T*=100 K, a minimal in the MR is seen at *H*~5.5 T. This unusual *H*-induced MR minimal is followed by evident hysteresis, indicating that the system undergoes a metastable state when sweeping *H*. Similar MR minimal was recently seen in heavy fermion metal CeIn$_3$ with large SOC, while the critical field (~60 T) was found to be one order of magnitude higher than the present case [27].

## 2.2. AMR contour rotation

To gain further insight into the field induced anisotropic transport, comprehensive AMR characterizations have been carried out, as shown in **Figure 3**. The device geometry for the AMR measurements is shown in Figure 3(a), in which the exciting current *I* is applied always along the [100] direction, *H* is rotated in-plane, and *Φ* is defined as the angle between *H* and *I*. Fourfold AMR =[*R*(*Φ*)-*R*(0)]/*R*(0) effect is generally seen below $T_C$ in the thin films, shown in Figure 3(b)-(c). The fourfold AMR symmetry excludes a normal AMR origin which simply depends on the relative angle between the exciting current and magnetic field with a relationship of *R*(*Φ*)~sin*Φ*, but should be mainly attributed to the magnetocrystalline anisotropy which depends on the relative angle between spin-axis and crystal-axis. This is further confirmed by a direct correspondence between the AMR symmetry and the tetragonal structure of $Sr_2IrO_4$.

An obvious effect seen in Figure 3(b) is that increasing the magnetic field up to *H*=7 T causes an in-plane AMR contour rotation by about 45°. For instance, the pristine fourfold AMR minima are changed to be maximal positions in the newly developed fourfold AMR. While the AMR contour rotation can be seen clearly, the AMR at 9 T may still be mixed with little low-field AMR component, indicating a much higher field that is desired to completely suppress the low field characteristic. This AMR contour rotation can still be seen at $T$=150 K and at a bit lower field $H$=5 T, shown in Figure 3(c). Such AMR contour rotation is consistent with the intercross behavior seen in the MR curves (Figure 2(c) and (d)). To distinguish the two types of fourfold AMR appearing before and after rotation, they are denoted as ARM-I and AMR-II, respectively.

Figure 3(d) presents complete AMR map at $T$=50 K, which were collected under external magnetic field ranging from $H$=0.1 T to 9 T. In prior to the onset of the weak FM phase at $H$<0.2 T, a twofold AMR is observed, which roughly follows a harmonic sin$\Phi$ dependence and can be ascribed to a standard non-crystalline AMR. As $H$>0.2 T, the weak FM phase emerges, which can be utilized to drive the AFM-axis travelling basal plane of $Sr_2IrO_4$. As a consequence, the four-fold AMR-I arises. A closer checking identifies minima of the AMR-I at $\Phi$ ~50°+$n\pi$/2 ($n$=0, 1, 2, 3). Note that the magnetic easy axis of $Sr_2IrO_4$ is along the [110] direction ($\Phi$~45°, not exactly because of the isospin canting) [17, 18]. Therefore, the fourfold AMR-I can be understood by the scheme of magnetocrystalline anisotropy, while a spot of contribution from normal AMR effect cannot be excluded considering the planar device geometry.

Further increasing $H$ to ~5 T causes significant suppression of the AMR-I, and instead activates another fourfold AMR symmetry (i.e., the AMR-II, in which the peaks are indicated by arrows) as $H$ >5 T, which has a ~45° shift relative to the AMR-I. A critical field $H_c$ of the AMR-I to AMR-II transition is estimated to be $H=H_c$~5 T, by plotting the peak positions of the AMR as a function of $H$, shown in Figure 3(e). The peak positions can be more precisely determined but don't change the symmetry of the AMR. Similar AMR transition can be seen in various temperatures below $T_C$ (**Figure S4**), and the critical field $H_c$ shows gradual decrease with increasing $T$, shown in Figure 3(f). By extrapolating the $H_c$-$T$ curve, a critical field of $H_c$ ~10 T can be obtained at $T$=0 K. Regarding the AMR-II, it still has a repetition duration of ~90°, while its minima appear along the pseudo-tetragonal axes of $Sr_2IrO_4$ (i.e., the [100] direction) which have a 45° shift relative to the basal magnetic easy axes (i.e., the [110] direction). In this sense, the emergence of the AMR-II could not be explained simply by the magnetocrystalline anisotropy as for the AMR-I. This will be discussed in details by combining with first principles calculations in the discussion section.

Stable directions of the antiferromagnetically ordered spins are separated by the magnetocrystalline anisotropy energy, and switching in-between these states by overcoming the energy barrier may lead to hysteresis. On one side, studying the switching dynamics would be of benefit to deeply understand the observed AMR effects. On the other side, for memory applications, hysteresis allows for non-volatile recording, while non-hysteretic may provide low-loss in magnetic sensors [6, 28]. To address this, **Figure 4**(a) presents AMR traces recorded by rotating $H$ from $\Phi=0°$ to $\Phi=360°$, and then back to $\Phi=0°$ at $T=50$ K. Note that there is no hysteresis at the rotation starting position, which can be used to disregard buckling problems in our experiments. This is further confirmed by repeating the measurements with different initial angles. Evident hysteresis can be seen in the AMR traces especially at low field range. For instance, the hysteresis is as large as $\Delta\Phi\sim14°$ at $H=0.3$ T. Increasing $H$ causes notable suppression in $\Delta\Phi$, and finally a rather small $\Delta\Phi\sim1°$ is seen at $H=9$ T. Figure 4(b) shows $\Delta\Phi$ as a function of $H$, in which three regions can be identified. For $H<0.5$ T, $\Delta\Phi$ decreases rapidly from $\sim14°$ to $\sim5°$ upon increasing $H$. For 0.5 T$<H<$ 5 T, $\Delta\Phi$ enters a plateau without evident change. Further increasing $H$ causes a step-like sudden decrease in $\Delta\Phi(H)$ at $H\sim5$ T, and then $\Delta\Phi$ again evolves with $H$ steadily at a level of $\Delta\Phi\sim1°$ till $H=9$ T.

By comparing the AMR transition characteristics $\Phi(H)$ (Fig. 3(e)) and the hysteresis evolution $\Delta\Phi(H)$ (Figure 4(d)), we can see that the two show intimate correlation with each other. At low field region with $H<0.5$ T, the fourfold AMR-I is gradually stabilized, and the hysteresis $\Delta\Phi$ decreases from 14° to 5° with increasing $H$. The relatively large $\Delta\Phi$ may be due to additional contributions of stabilizing the weak FM phase. At 0.5 T$<H<$ 5 T, the AMR-I possesses uniform hysteresis with a certain $\Delta\Phi\sim5°$, arising mainly from the magnetocrystalline anisotropy energy in $Sr_2IrO_4$ as discussed above. At 5 T$<H<$ 9 T, however, the AMR-II shows a nearly non-hysteretic behavior with minor $\Delta\Phi\sim1°$, suggesting that the isospins could go freely through the crystal axes in the thin films.

### 2.3. Theory of AMR of $Sr_2IrO_4$

We now have shown the magnetic field induced fourfold AMR contour rotation and concurrent MR minimal in the spin-orbit AFM SIO/STO thin films. These phenomena have not been reported yet, to the best of our knowledge.

To understand the microscopic physics of the unusual anisotropic magneto-transport in the thin films, we performed calculations based on density functional theory (DFT) using Vienna *ab initio* Simulation Package (VASP). First, the bulk $Sr_2IrO_4$ is checked. The magnetic ground state of $Sr_2IrO_4$ is confirmed to be the basal AFM arrangement with Ir isospins pointing along the [110] direction (magnetic easy axis). The local moment of Ir is $\mu$~0.498 $\mu_B$ with orbital moment $\mu_L$~0.341 $\mu_B$ and spin moment $\mu_S$~0.157 $\mu_B$, giving rise to a ratio of $\mu_L/\mu_S$~2.17. Moreover, the calculated canting angle of Ir moment is $\phi$~±10.7° with respect to the [110] direction. These results are well agreeing with previous experimental and theoretical results [17, 21, 22, 29].

We then perform the calculations for strained $Sr_2IrO_4$ to simulate thin films on the top of (001) STO substrate. The calculated results are listed in **Table I**. After including the strain effect, the magnetic easy axis remains the [110] direction. The ratio of $\mu_L/\mu_S$ is found to be ~2.58 for SIO/STO, although the total magnetic moment of Ir is almost unchanged (<5%) upon strain. The isospin canting angle is meanwhile derived to be $\phi$~5.3° with respect to the [110] direction for SIO/STO. Physically, strain can modify Ir-O-Ir bond angles in the *ab* plane, which tune the single-ion anisotropy, Dzyaloshinskii-Moriya interaction, as well as exchanges. Thus, the canting angles are naturally changed a little bit upon strain. The evolutions of both $\mu_L/\mu_S$ ratio and $\phi$ with strain are in agreement with previous dynamical mean field theory calculations (DMFT) [29].

As shown in Table I, the calculated magnetocrystalline anisotropy energy is $\Delta E$ ~ 1.01 meV/u.c. (or $\Delta E$ ~0.13 meV/Ir). This is quite close to the calculated value ($\Delta E$ ~0.19 meV/Ir) for $Ba_2IrO_4$ [30]. Such magnetocrystalline anisotropy can explain the AMR-I observed under relatively low magnetic field. As in many other magnetic systems, relatively stronger suppression of magnetic scattering (a consequence larger MR) would be expected when magnetic field is applied along the magnetic easy axis, which is confirmed by the results shown in Fig. 2(c) and (d). Then the AMR-I can be straightforwardly understood using this conventional mechanism of axis-dependent suppression of magnetic scattering.

Regarding the AMR-II, the relatively more conductive channel is shifted by ~45° to along the [100] direction, although the [110] direction is the magnetic easy axis in $Sr_2IrO_4$. As aforementioned, an AMR-I to AMR-II transition field is derived to be $H_c$~10 T at $T$=0 K, which corresponds to a Zeeman energy difference 0.12-0.2 meV/Ir if taking the Ir moment as 0.2-0.4 $\mu_B$/Ir in $Sr_2IrO_4$ [17, 22]. Therefore, $H_c$ represents a critical field to overcome the magnetocrystalline anisotropy energy barrier. Thus, when $H>H_c$, all the isospins should fully rotate accompanying the external magnetic field.

We further calculated band structures for cases with the Ir isospins pointing along the [110] and the [100] directions, as shown in **Figure S5**. Similar to previous theoretical works [14, 29, 31], we found that taking the strong SOC and effective Hubbard repulsion $U_{eff}$ into account is vital for the gap-opening in Sr$_2$IrO$_4$. A smaller band-gap ($E_g^{[100]}$~25.9 meV) is revealed as the Ir isospins are aligned along the [100] direction, as compared with the [110] direction ($E_g^{[110]}$~35.0 meV). It should be noted that the gap calculated using the GGA-PBE in VASP is typically smaller than the experimental one (or other theoretical values calculated using DMFT) [14, 29]. Even though, these values are qualitatively comparable, i.e. $E_g^{[110]}>E_g^{[100]}$, which is in agreement with the previous calculated results [11]. This difference can well explain the emergence of the AMR-II in the SIO/STO thin films. The band gap $E_g^{[100]}$ (~25.9 meV) is always smaller than $E_g^{[110]}$ (~35.0 meV). The isospin's direction can rotate with increasing magnetic field, not $E_g$'s. Under small magnetic fields, the energy is lower when isospins pointing along the [110] direction and the MR is larger along the [110], considering the magnetic easy axis. Under high magnetic fields, the weak FM magnetization are nearly saturated no matter the magnetic field is applied along the [100] or [110] direction. Thus the AMR-I, which is contributed by the suppression of magnetic scattering, should be negligible in this situation. Instead, since all isospins follow the magnetic field, the intrinsic band gaps tuned by isospins' direction will determine the transport, leading to the AMR-II.

According to the calculated band structures, the minimal in the MR of $H$//[110] can be explained as following. Magnetic twin domains have been demonstrated in bulk SIO, owing to the tetragonal symmetry [22]. In the present SIO/STO thin films, similar twinned magnetic domains are to be expected, since the tetragonal structure is preserved and the strain in the thin films is tiny, as evidenced by the structural characterizations. Upon increasing $H$, the domains are eventually aligned to let the FM moment approach $H$, leading to suppression of spin-scattering such as quick enhancement in MR at low fields (Fig. 2(c) and (d)). As $H \geq H_c$, the field is sufficient to overcome the anisotropy energy barrier, and thus the Ir isospins can pass the [100] direction (with smaller band-gap) and reach the final stable direction such as the [110] direction (with larger band-gap). Therefore, a MR minimal appears at $H_c$ in the MR with $H$//[110]. A sketch is shown in **Figure 5**. This picture is supported by the fact that the fields of MR minima at various temperatures well follows the relationship of $H_c(T)$ shown in Figure 3(f).

**3. Conclusion**

In conclusion, well controllable AMR effect and concurrent MR minimal are evidenced in the Sr$_2$IrO$_4$ thin films hosting a novel $J_{eff}$=1/2 Mott state, which can be observed in a broad temperature range up to the AFM transition. Our first-principles calculations reveal that these two phenomena should be mainly attributed to band structure engineering when rotating the AFM-axis laterally in the thin films. Our results unequivocally link the $J_{eff}$=1/2 Mott state to the AFM-based AMR, which would be scientifically interesting and important for AFM spintronics, since the realization of AMR represents an important step towards the manipulation and detection of AFM orders. In addition, our work evidences a direct correlation between the inherent AFM-lattice and the band structure in Sr$_2$IrO$_4$, which provides useful information to understand the nature of magnetic interactions in iridates.

## 4. Experimental section

The Sr$_2$IrO$_4$ thin films were epitaxially grown on STO (001) substrate using pulsed laser deposition system equipped with *in-situ* RHEED. The growth parameters were carefully optimized, and the details can be found in our previous work [26]. The film thickness of 40-nm was monitored by the RHEED intensity oscillations.

X-ray diffraction characterizations, including regular theta-2theta scan, reciprocal space mapping, and rocking curves were carried out using a Philips X'Pert diffractometer. Magnetic characterizations were performed using a superconducting quantum interference device by Quantum Design. The *M(T)* curves were measured for both field cooling and zero field cooling sequences, and the cooling and measuring field was set at *H*=0.1 T. Magnetization as a function of *H* were measured at various temperatures after a ZFC sequence. During the magnetization measurements, *H* was applied along the [100] direction. Electric transport measurements with exciting current *I* //[100] direction were performed using a standard four-probe method in a Quantum Design physical property measurement system equipped with a rotator module. With regard to the anisotropic magnetoresistance measurements, *I* is applied always along the [100] direction, and the magnetic field *H* is rotated within the basal plane of the films. Maximum magnetic field allowed by the transport measurement set-up is 9 T.

In DFT calculations, the Hubbard repulsion $U_{eff}$=$U$-$J$=3 eV is concluded and the SOC effect is considered with noncollinear spins. The plane-wave cutoff is 550 eV and the 6×6×2 Monkhorst-Pack *k*-points mesh is centered at $\Gamma$ points. Starting from the experimental tetragonal structure of bulk SIO, the lattice parameters and inner atomic positions are fully optimized till the Hellman-Feyman forces are all less than 0.01 eV/A [32]. To simulate the strain in thin films, the in-plane lattice constants of SIO are fixed to be the same as the substrates. Our calculations were done without considering surface and interface, noting that the films are already 40 nm in thickness, sufficiently thick that the bulk properties are believed to be dominant. The isospin moments are initialized along particular axes (without canting), but not artificially constrained. The magnetic caning is obtained via self-consistent calculations.

**Supporting Information:** Supporting Information is available from thee Wiley Online Library or from the authors.

**Acknowledgements:** We acknowledge fruitful discussions with Prof. Marin Alexe, Dr. Xavi Marti, Dr. Ignasi Fina, and Prof. Dietrich Hesse. This work was supported by the National Nature Science Foundation of China (Grant Nos. 11774106, 11674055, 51332006, 51721001, and 51431006), the National Key Research Projects of China (Grant No. 2016YFA0300101).

Received: ((will be filled in by the editorial staff))
Revised: ((will be filled in by the editorial staff))
Published online: ((will be filled in by the editorial staff))

**Figure captions:**

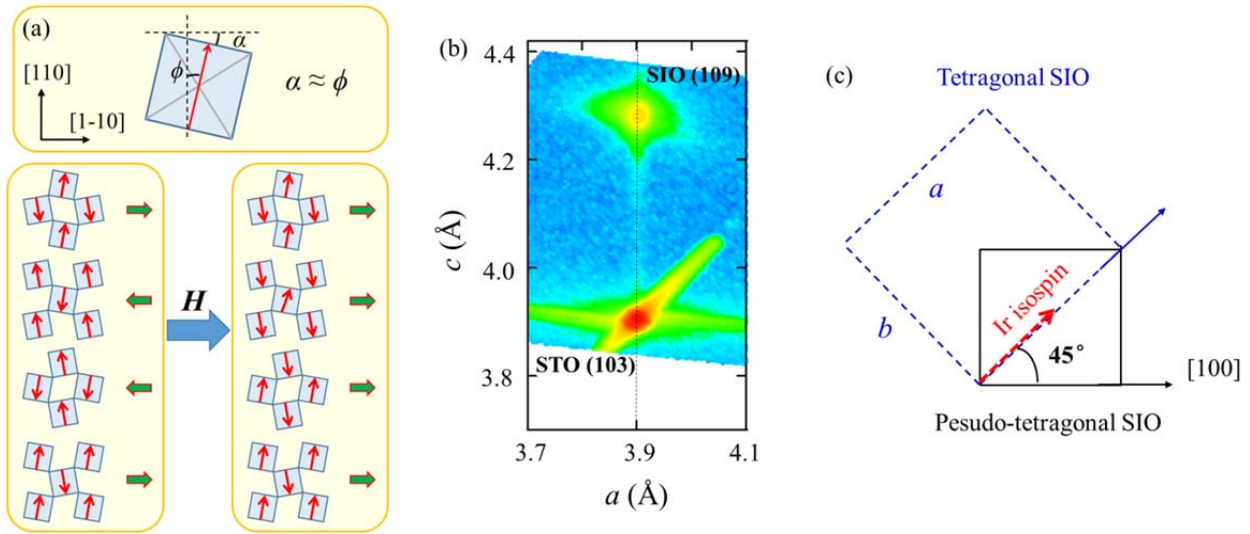

Figure 1: (a) The top panel shows the locking effect between octahedral rotation $\alpha$ and isospin (red arrow) canting $\phi$. The bottom panel shows the planar antiferromagnetic structure in $Sr_2IrO_4$. The olive arrows denote the net moment in each $IrO_2$ layer, arising from the isospin canting. The application of magnetic field in-plane would induce an isospin-flip transition, leading to a weak ferromagnetic phase. (b) Reciprocal space mapping around the (109) plane of the $Sr_2IrO_4$ thin films grown on (001) $SrTiO_3$ substrates. (c) Schematic of the basal planes of tetragonal (blue square) and psudotetragonal (black square) $Sr_2IrO_4$.

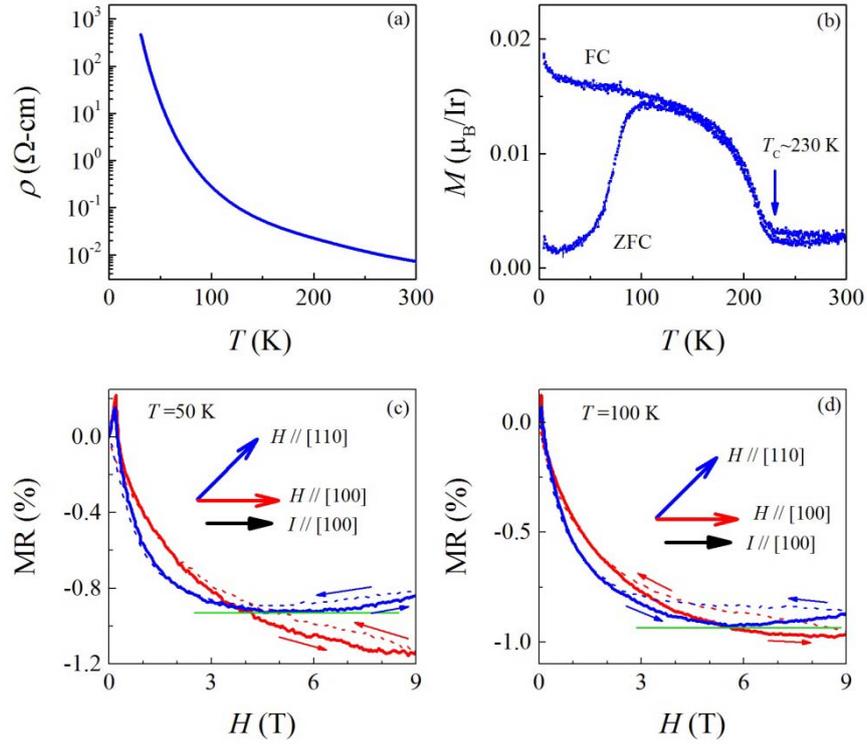

Figure 2: (a) Resistivity and (b) magnetization as a function of temperature. The $T_C \sim 230$ K is indicated by an arrow in (b). (c) and (d) show magnetoresistance of the SIO thin films measured with field applied along different directions at $T=50$ K, and $T=100$ K, respectively. The inset shows a sketchy of the measurements. Solid (dashed) curves were obtained by sweeping up (down) $H$. The minima in MR curves with $H//[110]$ are highlighted by green lines.

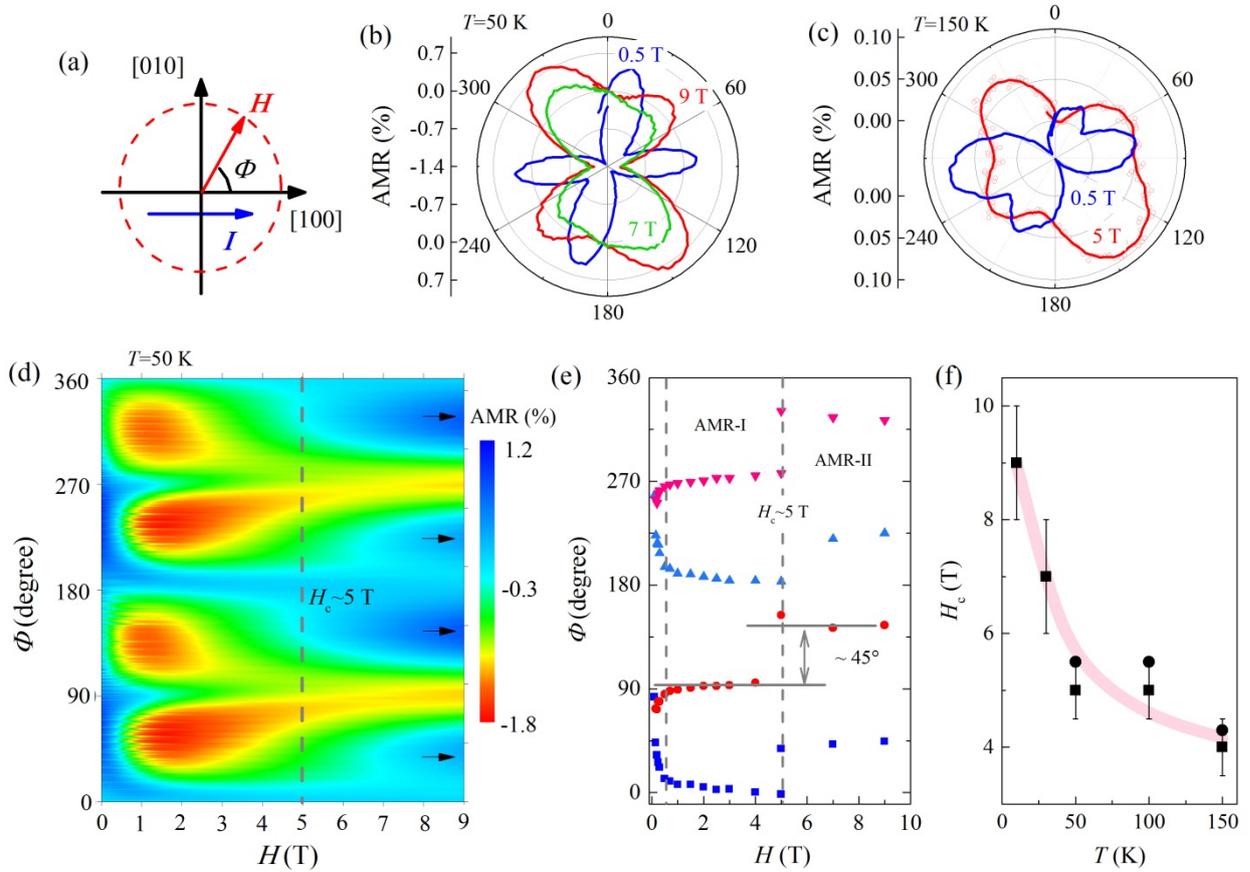

Figure 3: (a) Schematic of the anisotropic magnetoresistance measurement, in which the current is applied along the [100] direction, and magnetic field is rotated in-plane. $\Phi$ is the angle between magnetic field $H$ and the current $I$. (b) and (c) show the anisotropic magnetoresistance measured under various magnetic field at $T$=50 K and $T$=150 K for the thin films, respectively. (d) Anisotropic magnetoresistance map collected under a series of magnetic fields at $T$=50 K for the thin films. The high-field AMR peaks are indicated by arrows. (e) Peak positions of the AMR curves in (d) as a function of $H$, in which the contour rotation is clearly seen at $H_c$~5 T. AMR-I and AMR-II represent the AMR effect that appearing before and after the rotation. The relative shift between AMR-I and AMR-II is ~45°. (f) The critical field $H_c$ (squares) of the AMR-I to AMR-II transition as a function of temperature. The dots represent fields of the MR minima at various temperatures.

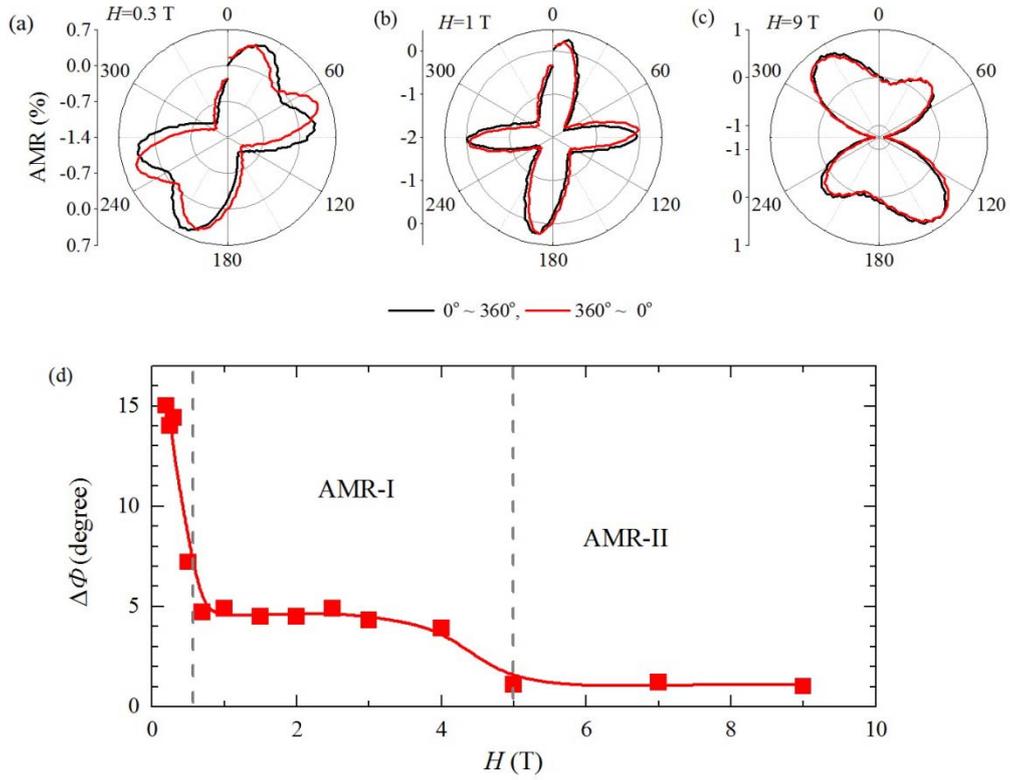

Figure 4: Hysteretic AMR traces measured under various magnetic fields (a) $H$=0.3 T, (b) $H$=1 T, and (c) $H$=9 T at $T$=50 K in SIO/STO thin films. The magnetic field is rotated from $\Phi$=0° to 360° (black curves) and then backwards from 360° to 0° (red curves). (d) The derived hysteresis size $\Delta\Phi$ as a function of $H$, in which step-like features are seen corresponding to the AMR-I and AMR-II.

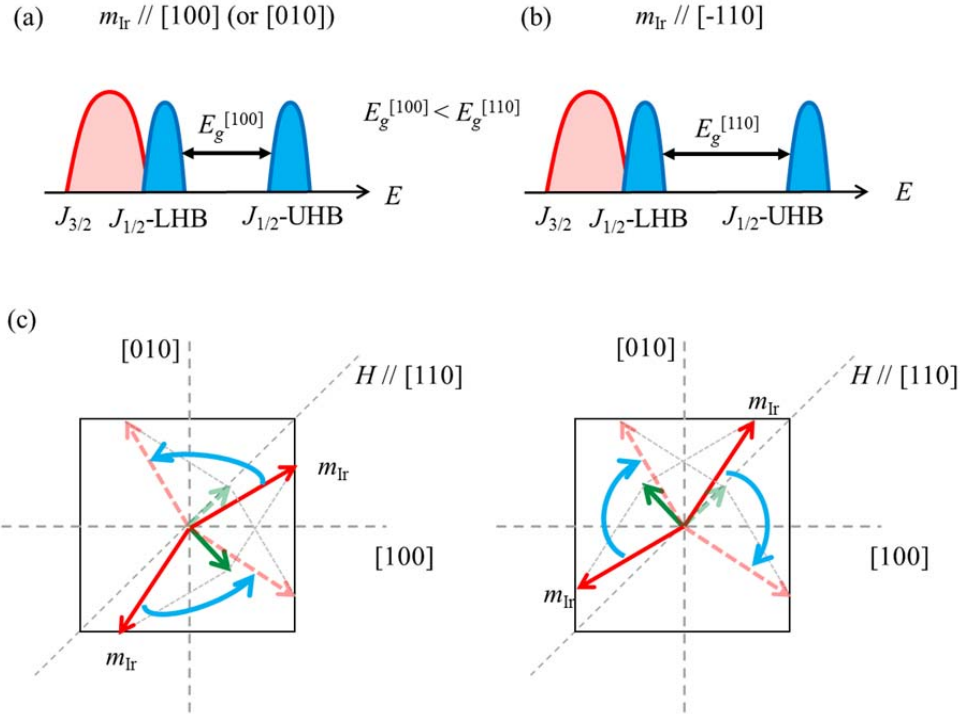

Figure 5: (a) and (b) schematically show the band structures as Ir isospin is aligned along the [100] and [110] directions, respectively. (c) Sketch of $H$ driven rotation of the Ir isospins $m_{Ir}$ (red arrows) as $H>H_c$. Olive arrows represent the FM moments due to isospin-canting. Solid (dashed) arrows represent the initial (final) positions of the moments.

**Table I.** The energy difference for a minimal unit cell (eight formula units), local spin moment ($\mu_S$) within the default Wigner-Seitz sphere, and orbital moment ($\mu_L$) for each Ir of $Sr_2IrO_4$. The item of "isospin angle" means a canting angle relative to the initialized direction of the isospin.

| Initial moments direction | | [110] | [100] |
|---|---|---|---|
| Bulk SIO | Energy (meV) | 0 | 1.24 |
| | $\mu=\mu_S+\mu_L$ ($\mu_B$/Ir) | 0.498 | 0.496 |
| | isospin angle (°) | ±10.7 | +9.6/-11.6 |
| | $\mu_S$ ($\mu_B$) | 0.157 | 0.155 |
| | $\mu_L$ ($\mu_B$) | 0.341 | 0.341 |
| SIO/STO (001) | Energy (meV) | 0 | 1.01 |
| | $\mu=\mu_S+\mu_L$ ($\mu_B$/Ir) | 0.476 | 0.473 |
| | isospin angle (°) | ±5.3 | +1.1/-8.7 |
| | $\mu_S$ ($\mu_B$) | 0.133 | 0.131 |
| | $\mu_L$ ($\mu_B$) | 0.343 | 0.342 |
| | $E_g$ (meV) | 35.0 | 25.9 |